# Exciton transfer dynamics and quantum diffusion in a lattice of two level systems: Interplay between transport and coherent population transfer dynamics

**Rajesh Dutta and Biman Bagchi***

*E-mail: bbagchi@sscu.iisc.ernet.in, profbiman@gmail.com

Solid State and Structural Chemistry Unit, Indian Institute of Science, Bangalore-560012, India.

## *ABSTRACT*

**We study excitation transfer dynamics in a lattice of two level systems characterized by dynamic disorder. The diagonal and off-diagonal energy disorders arise from the coupling of system and bath. We consider both the same and the independent bath limits. In case of independent bath all diagonal and off-diagonal bath coupling elements fluctuate independently of each other and the dynamics is complicated. We obtain the time dependent population distribution by solving Kubo's quantum stochastic Liouville equation (QSLE). The main result of our study is both the population transfer dynamics and the mean square displacement of the exciton behave the similar way in the same and independent bath cases in the Markovian limit. However, these two baths can give rise to markedly different behavior in the non-Markovian limit where coherent transport becomes important. *We note that previously only the same bath case has been studied in the non-Markovian limit*. There are also several additional new results as follows. (i) Exciton migration remains coherent all the time for an average, non-zero off-diagonal coupling**

**value *J* for the same bath case while it becomes incoherent for independent bath case in the Markovian limit. (ii) An oscillatory behavior of the population transfer dynamics supports the coherent mode of transfer of exciton. (iii) Agreement with available analytical expression of mean squared displacement is good in the Markovian limit for independent bath case with off-diagonal fluctuation but only qualitative in the non-Markovian limit for which no complete analytical solution is available. (iv) Transition from coherent to incoherent transport is observed in the independent bath case with diagonal fluctuation when the bath is made progressively more Markovian. An analytical study that shows coherence is propagated through excited bath states is also provided. (v) The correlation time of the bath plays a unique role in dictating the diffusive spread that is not anticipated in a Markovian treatment.**

## I. INTRODUCTION

Diffusion of a tagged particle in a classical random system is well-understood in terms of a random walk of the particle in a random dynamic environment. For example, in the liquid state, one can obtain an accurate estimate of the diffusion coefficient in terms of Stokes-Einstein relation.[1-5] Another interesting case is the diffusion of a Brownian particle that migrates in the presence of a quenched disorder, like a random barrier.[6,7] However, the situation is different for quantum systems where the concept of friction is not well defined. Diffusion of a quantum particle is complicated by the presence of coherences that extend the size of the quantum entity (particle, hole, energy) that diffuses. There are also important issues like localization due to randomness and the impact of purely quantum effects such as tunneling. Despite considerable work, study of quantum diffusion remained at its infancy.

Much of the discussions on transport in quantum systems are focused on localization of a quantum particle due to static randomness (that is, quenched disorder), as in Anderson localization.[8-11] There are several studies and application of dynamical disorder[12-16] in classical system, however, quantum transport in the presence of a dynamic disorder has been less discussed. Dynamic disorder for quantum system has been used to describe line shape calculation,[17,18] vibrational energy relaxation[19] and more recently to explore excitation energy transfer in photosynthesis as well as model systems (study on this field is provided in later part of this section).[20,21] In case of dynamic disorder for quantum system, the localization may absent yet the quantum nature dominates migration, and the situation is quite different from the classical case. In this limit, there could be situations where diffusion coefficient might not exist in the sense that mean square displacement (MSD) does not increase linearly with time $t$ even in the long time limit. Because of coherences, the growth of MSD can be faster than $t$. Several diffusion processes are known where MSD can either be sub-diffusive or super-diffusive.[22]

An important example of quantum diffusion is provided by the migration of an initially localized excitation in a linear array of equally spaced two level systems. Migration is affected by off-diagonals terms in the interaction Hamiltonian. Since the bath fluctuates with time, the interaction is stochastic.

Despite the simplification of the model Hamiltonian also known as Haken-Strobl-Reineker (HSR) Hamiltonian, that is often employed in the analysis of migration of exciton, many aspects of this model remain unsettled. The most important lacuna is the unavailability of an analytical expression for the general case of the system interaction with a non-Markovian bath. In the case of a non-Markovian bath, the importance of coherent transport can become enhanced to a great extent.

The initial theories of Haken-Strobl-Reineker[23-25] and of Silbey *et al.*[26-28] are based on the following standard Hamiltonian

$$H_{tot} = H_S + H_B + H_{\text{int}} \tag{1}$$

where system (exciton) Hamiltonian is defined as

$$H_S = \sum_k E_0 |k\rangle\langle k| + \sum_{\substack{k,l \\ k \neq l}} J_{kl} |k\rangle\langle l| \quad . \tag{2}$$

where $E_0$ is the energy of an electronic exciton localized at site *k* and $J_{kl}$ is the time independent off-diagonal interaction between excitations at site *k* and $l$ . $H_B$ is the bath Hamiltonian which may be due to the phonon contribution and $H_{\text{int}}$ is the interaction Hamiltonian between the system (exciton) and the bath. In this work we consider the total Hamiltonian in the interaction representation of the bath Hamiltonian, so that the interaction $V$ is time dependent and can be modeled by a stochastic function with known statistical properties. In this interaction representation, the Hamiltonian can be represented as follows,

$$H(t) = H_S + V(t) \tag{3}$$

where, setting $\hbar = 1$, $V(t)$ has the following form

$$V(t) = e^{iH_B t} V e^{-iH_B t} . \tag{4}$$

Thus in the interaction representation, the coupling potential is time dependent which we write as

$$V(t) = \sum_k |k\rangle\langle k| V_d(t) + \sum_{\substack{k,l \\ k \neq l}} |k\rangle\langle l| V_{od}(t) \quad . \tag{5}$$

Here $V_d(t)$ and $V_{od}(t)$ denote diagonal (local) and off-diagonal (non-local) parts of the fluctuating potential $V_{int}$. We assume both the fluctuation to be described by Poisson stochastic process, although the general formalism is valid for other processes also. In fact, the Poisson bath can be considered as a limiting form of Gaussian bath.[19] For Poisson bath case, diagonal and off-diagonal matrix elements jump between two values (say $\pm V$) such that the average of each matrix element will be zero. In this case we consider two cases as (i) same bath and (ii) independent bath case (details are provided in the forthcoming section).

Several decades ago, Haken, Strobl and Reineker,[23-25] and independently Silbey and co-workers[26-28] studied the Markovian limit for infinite system. Even in the Markovian limit of bath fluctuations, this model gives rise to non-trivial prediction. First, the total diffusion constant is expressed as

$$D_{HR} = a^2\left[2\gamma_{od} + \frac{J^2}{\gamma_d + 2\gamma_{od}}\right] \tag{6}$$

where $\gamma_d$ and $\gamma_{od}$ describe the effective rates of the diagonal and off-diagonal fluctuations.

In the Markovian limit $\gamma_d$ and $\gamma_{od}$ both are related to fluctuation strength ($V_d$ and $V_{od}$) and the respective correlation times ($b_d$ and $b_{od}$) in the following (well-known) fashion

$$\underset{V_d, b_d \to \infty}{lt} \frac{V_d^2}{b_d} = \gamma_d$$

$$\underset{V_{od}, b_{od} \to \infty}{lt} \frac{V_{od}^2}{b_{od}} = \gamma_{od}\,. \tag{7}$$

This manner of expressing rate constants in terms of coupling constants and correlation times, and their limits is well-known. Later we provide comparison of the Haken-Reineker diffusion coefficient with our model.

The main limitation of many of the previous works was the consideration in the Gaussian white noise limit. Bagchi and Oxtoby[29] discussed the limitation of these theories, using Kubo's quantum stochastic Liouville equation.[30] We shall return to this point later.

On the experimental side, Zewail and Harris,[31-33] in an elegant study, observed that a coherent state in a dimer can last for $\sim 10^4$ exchange time between the dimeric states before the scattering of the phonon environment destroy the coherence or wave like motion of the exciton. They performed very low temperature (～1.7°K) optically detected magnetic resonance experiments in zero magnetic field on dimers of 1, 2,4,5 tetra-chloro-benzene embedded in the deuterated compound. At the similar time Harris and Fayer[34,35] also carried out a series of experiments on triplet exciton in 1,2,4,5 tetra-chloro-benzene and observed coherent migration of exciton at very low temperature.

Excitation energy transfer has paramount importance in case of conjugated polymers. Conjugate polymers can be used in constructing light emitting diodes, photovoltaic cells and optical sensors as well as several others opto-electronic devices. The main interest in this field arises as the performance of these devices depends upon the transport of excitations in polymer chain. Collini and Scholes[36,37] have observed coherent intrachain excitation energy transfer in polymer at room temperature. Saini and Bagchi[38] have studied energy transfer in conjugated polymers using FRET theory and observed that the excitation energy is photochemically funneled from smaller to larger chromophores upon excitation. Burghardt and co-workers[39,40] have explained electronic coherence is preserved at shortest time scale and decoherence settles in longer time scales using non-Markovian system-environment dynamics by a hierarchical series of approximate spectral densities and also studied coherent excitations transfer starting from partially delocalized exciton state using model Hamiltonian for poly-phenylen-vinylene (PPV) type systems. Barford, Bittner and ward[41] have provided a general definition of absorbing and emitting chromophore and applied the theory to calculate exciton diffusion length in PPV. Bittner and co-workers[42] have observed the role quantum coherence and energy fluctuation in the creation of charge-separated states i.e. fission of exciton in polymeric type II hetero junction devices. Recently Tozer and Barford[43] have investigated intra-chain exciton dynamics of a polymer in solution induced by the Brownian rotational motion of the monomers.

In another important application of excitation energy transfer, it has been observed that the efficiency of energy transfer from chromophores to reaction center in photosynthetic system is near unity.[44,45] This type high efficiency in energy transfer suggests a possible way to make highly efficient artificial solar cell. But the main obstruction is due to lack of experimental and theoretical knowledge about the process. In photosynthetic systems chromophore are surrounded by proteins which play a crucial role in energy transfer process. Much of interests in this field arise because of presence of coherent energy transfer dynamics in noisy biological environment. Pioneering work by Fleming and co-workers[46] have revealed long lived quantum coherence in FMO protein (Fenna-Matthews-olson) using two-dimensional Fourier transform electronic spectroscopy. Later quantum coherence has been observed in one of the most essential photosynthetic complex LHCII[47]. Lee *et al.*[48] have explored coherent dynamics in the reaction center of purple bacteria using two-color electronic coherence photon echo-technique.

Several theories have been developed to describe energy transfer dynamics among molecules and chromophores in condensed matter systems. The most popular among them is the fluorescence (or, Förster) resonance energy transfer (FRET) theory[49-53] which is based on weak coupling approximation. The second approach employs the Redfield equation[54-56] which is based on Markovian approximation. Both the theories ignore coherence. Recently Ishizaki and Fleming[57,58] have used reduced hierarchy equation, which is based on path integral approach for quantum dissipative system developed by Kubo and Tanimura,[59] to study the energy transfer dynamics. Another well-known approach is based on polaron transfer technique developed by Jang and co-workers.[60] Recently Aspuru-Guzik and co-workers[61,62] have explained environment assisted quantum transport in real and model systems. Silbey and co-workers[63-65] have calculated efficiency of exciton transfer in the case of dimer model system and population relaxation for FMO complex. Few years ago Chen and Silbey[66] have considered independent dichotomic noise for dimer and FMO complex and used only diagonal fluctuation to calculate population relaxation and efficiency for dimer system and photosynthetic system respectively though the method of truncation of infinite cumulant expansion by Chen and Sibey is not fully correct. Most of the above studies have employed Markovian approximation which assumes either a short bath correlation time or weak coupling with the bath, as described by Eq. (7). The non-Markovian limit has not been studied in detail.

The main objectives of the present work are as follows:

(i) To study the exciton transfer dynamics for discrete model system in both Markovian and non-Markovian limits.

(ii) To perform both theoretical and numerical analyses of exciton transfer dynamics for the same and the independent bath cases. We have separately considered the diagonal and off-diagonal fluctuations for the independent bath case.

(iii) To calculate the population time correlation function (PTCF) and mean squared displacement (MSD) in non-Markovian and Markovian limits in order to observe coherent and incoherent transport and the transition from coherent to incoherent transport.

It is worthwhile to point out here that only the same bath limiting case has been studied previously in the non-Markovian limit. We obtain several potentially interesting results. (i) When the average off-diagonal coupling is nonzero, the exciton migration remains coherent (in the same bath limit) even at long times while it becomes incoherent in the independent bath case in Markovian limit. In contrast to the prediction of the Markovian theories, the bath correlation time plays an important role distinct from the fluctuation in coupling elements. (ii) Coherent energy transfer is manifested in the oscillatory behavior of the energy transfer dynamics accompanied by faster-than diffusive spread of the exciton from the original position. (iii) Fairly good agreement with available analytical expression of mean squared displacement is observed in the Markovian limit for independent bath case with off-diagonal fluctuation but only qualitative in the non-Markovian limit for which no complete analytical solution is available. (iv) We observe transition from coherent to incoherent transport in the independent bath case with diagonal fluctuation when the bath is made progressively more Markovian. We present an analytical study that shows coherence propagates through excited bath states. (v) The correlation time of the bath plays a unique role in dictating the diffusive spread that is not anticipated in a Markovian treatment.

The organization of the rest of the paper is as follows: In the next section we discuss Kubo's stochastic Liouville equation (QSLE). In **Sec. III** we explain rate equation description for energy transfer dynamics and in **Sec. IV**, we describe population time correlation function using QSLE and rate equation description. In **Sec. V**, we illustrate the mean square displacement. In **Sec. VI**, we explain the relation between coherence and diffusion. Subsequently, in **Sec. VII**, we elucidate the exciton transfer dynamics in long chain and in **Sec. VIII**, we discuss the connection between rate of energy transfer and diffusion coefficient. Finally, in **Sec. IX**, we draw the conclusion.

Derivation of coupled equation of motion for both same and independent bath case can be found in **APPENDIX A** and **APPENDIX B** respectively whereas coupled equation of motion for 7 site system from rate equation description is provided in **APPENDIX C.**

## II. QUANTUM STOCHASTIC LIOUVILLE EQUATION (QSLE) AND COUPLED EQUATION OF MOTION

Quantum stochastic Liouville equation (also the classical version of the equation) was developed by Kubo[30] to incorporate stochastic energy fluctuations in Liouville equation. The theory is ideally suited when the system is quantum but the bath is classical. It is a useful equation and has already been applied to study electron spin resonance (ESR) and nuclear magnetic resonance (NMR) studies as well as vibrational relaxation (for both energy and phase relaxation).

Study of such quantum system-classical bath has a long history.[67-70] Notable contribution to this area particularly in the context of spectroscopy[71,72] has been made by Skinner and co-workers. If the system is harmonic oscillator one can provide theoretical calculation comparable with experiments. Hernandez and Voth[73] explored the role of classical mechanics in defining coherence in quantum system through the calculation of time correlation function.

Bagchi and Oxtoby used the QSLE to explain exciton transport in one dimensional lattice considering same bath and independent bath for continuum model. In this study for the same bath case transport remains coherent for non-zero value of exchange integral (for both Markovian and non-Markovian limit) whereas for the independent bath case transport is incoherent in Markovian limit. Study of Haken and coworkers and also Silbey *et al.* have drawn the similar conclusion that exciton motion is diffusive for continuum model (Markovian theory).

The present work is an extension of the studies of Haken *et al.*, Silbey *et al.* and Bagchi-Oxtoby.

To obtain the QSLE, one has to start with the quantum Liouville equation and the formalism are provided as follows,

$$\frac{d\rho}{dt} = -\frac{i}{\hbar}\left[H(t), \rho\right]. \tag{8}$$

The master equation for probability measure can be expressed as,

$$\frac{dW(\mathrm{V},t)}{dt} = \Gamma_{\mathrm{V}} W(\mathrm{V},t). \tag{9}$$

where $\mathrm{V}$ is random variable, $W(\mathrm{V},t)$ is the probability measure and $\Gamma_{\mathrm{V}}$ is the stochastic diffusion operator.

Next we can define joint probability distribution $P\left(\mathrm{V},\rho,t\right)$ as follows

$$P\left(\mathrm{V},\rho,t\right) = \left\langle \delta\left(\mathrm{V}-\mathrm{V}(t)\right)\delta\left(\rho-\rho(t)\right)\right\rangle. \tag{10}$$

Reduced density matrix $\sigma(t)$ can be defined as,

$$\sigma(t) = \int d\rho \; \rho \; P\left(\mathrm{V},\rho,t\right). \tag{11}$$

We recombine the above equations to obtain

$$\frac{d\sigma}{dt} = -\frac{i}{\hbar}\left[H(t), \sigma\right] + \Gamma_{\mathrm{V}}\sigma. \tag{12}$$

This is Kubo's stochastic Liouville equation. In the subsequent steps, it is expanded in the eigenstates of the bath operator to obtain a system of coupled equations. The system of equations can be solved analytically in the continuum (in space) limit but usually solved numerically to obtain detailed behavior.

### 1. Coupled equation of motion for the same bath case

If diagonal and off-diagonal fluctuation arises due to the coupling with the same bath, the time dependent Hamiltonian only depends upon one random variable. We expand $\sigma$ in the eigenstates of the diffusion operator $\Gamma$ that describe both diagonal and off-diagonal fluctuation and substituting back into Eq. (12) (detailed derivation is provided in APPENDIX A) we obtain the following coupled equation of motion (we have considered $\hbar = 1$)

$$\frac{d\sigma_m}{dt} = -iH_{ex}^x \sigma_m - iV_d \sum_{m'=0}^{1} (\delta_{m+1,m'} + \delta_{m-1,m'}) \times \left( \sum_k |k\rangle\langle k| \right)^x \sigma_{m'}$$
$$- iV_{od} \sum_{m'=0}^{1} (\delta_{m+1,m'} + \delta_{m-1,m'}) \times \left( \sum_{\substack{k,l \\ k\neq l}} |k\rangle\langle l| \right)^x \sigma_{m'} - mb\sigma_m \tag{13}$$

where $O^x f = Of - fO$ . We have considered that units of all the parameters are same and designated as time inverse. For the same bath case the coupled equation of motion of exciton, Eq. (13) is solved numerically by the Runge-Kutta fourth order method with assuming initial condition i.e. the exciton initially placed at site 1. For same bath case population of each site can be denoted as follows,

$$P_n(t) = \langle n|\sigma_0|n\rangle \tag{14}$$

where $n$ is the site number.

**2. Coupled equation of motion for the independent bath case**

For the independent bath case the nearest neighbor diagonal and off-diagonal bath coupling elements are independent of each other i.e. they are uncorrelated with each other all times. Though all the diagonal fluctuations are independent, we assume that fluctuation strength $V_d$ and correlation time $b_d$ have same value. This is also true for off-diagonal fluctuation. The assumption can be relaxed without much difficulty, although the algebra gets more involved. The QSLE for the independent bath case can be written as

$$\frac{d\sigma}{dt} = -\frac{i}{\hbar}\left[H(t),\sigma\right] + \sum_j \Gamma_j \sigma \quad . \tag{15}$$

where $\Gamma_j$ is the stochastic force for $j^{\text{th}}$ random force.

Following the same procedure that we have already discussed for same bath case one can obtain coupled equation of motion of exciton for the independent bath case.

Let us consider a two sites model where two independent diagonal and one off-diagonal fluctuation are available. Hence one can write the coupled equation of motion for this system as follows (we have considered $\hbar = 1$) (derivation is provided in APPENDIX B)

$$\begin{aligned}\frac{d\sigma_{jkl}}{dt} = & -iE_0\left(|1\rangle\langle 1| + |2\rangle\langle 2|\right)^x \sigma_{jkl} - iJ\left(|1\rangle\langle 2| + |2\rangle\langle 1|\right)^x \sigma_{jkl} - iV_d \sum_{j'=0}^{1}\left(\delta_{j+1,j'} + \delta_{j-1,j'}\right)\left(|1\rangle\langle 1|\right)^x \sigma_{j'kl} \\ & - iV_d \sum_{k'=0}^{1}\left(\delta_{k+1,k'} + \delta_{k-1,k'}\right)\left(|2\rangle\langle 2|\right)^x \sigma_{jk'l} - iV_{od} \sum_{l'=0}^{1}\left(\delta_{l+1,l'} + \delta_{l-1,l'}\right)\left(|1\rangle\langle 2| + |2\rangle\langle 1|\right)^x \sigma_{jkl'} \\ & - jb_d \sigma_{jkl} - kb_d \sigma_{jkl} - lb_{od}\sigma_{jkl} \, .\end{aligned} \tag{16}$$

For the independent bath case coupled equation of motion of exciton for 3, 5, 7 and 9 site model is solved using Eq. (15) (final equation is similar but the extrapolated version of Eq. (16)). In this case number of coupled differential equations is huge and we have solved numerically using Runge-Kutta fourth order method. In this study we have assumed exciton is initially placed at site 1 and considered diagonal as well as off-diagonal fluctuation separately. For independent bath case population for each site can be represented as follows,

$$P_n(t) = \langle n|\sigma_{\prod_{i=1}^{2N-1} a_i}|n\rangle \tag{17}$$

where $n$ is the site number, $N$ is total site number and $a_1, a_2, a_3, \ldots a_{2N-1}$ all the elements are zero. For $N$ sites model total number of diagonal and off-diagonal elements is $2N-1$.

The coherence is propagated through the excited state bath mode $\sigma_1(t)$ for both same and independent bath case. If the diagonal and off-diagonal term of reduced density matrix in excited state bath mode is zero (or negligible), the exciton motion quickly attains diffusive behavior.

## III. RATE EQUATION DESCRIPTION AND COUPLED EQUATION OF MOTION

Since the sites are identical and the noise is homogeneous and uniform, we can assign a constant rate of transmission from one site to another. Later we shall derive expressions of the rate in terms of the stochastic parameters *V* and *b*. This constant rate allows us to write down a set of master equations that easily provide a rate equation description of the population transfer dynamics. In essence, this rate equation description consists of coupled equation of motion for population or energy transfer. The coupled equation of motions for population or energy transfer dynamics in 7 sites system are provided in the **APPENDIX C**. We have calculated population relaxation i.e. $P_n(t)$ where $n$ denotes the site number. We have numerically calculated the population of each site using Runge-Kutta fourth order method. We have assumed that only site 1 is initially populated. One can obtain the mean squared displacement for this type of discrete systems using the following formula

$$\left\langle n^2 \right\rangle = \sum_n n^2 P_n(t) \tag{18}$$

where $n$ is the site number. One can calculate diffusion coefficient from the mean squared displacement.

## IV. POPULATION TIME CORRELATION FUNCTION (PTCF)

We have defined population time correlation function as follows,

$$C_n^P(t) = \frac{P_n(t) - P_n(\infty)}{P_n(0) - P_n(\infty)} \tag{19}$$

where $P_n$ is the population of $n^{\text{th}}$ site. We have calculated PTCF for both same and independent bath for 7 sites model. For independent bath case with only off-diagonal fluctuation, population for 7 sites model can be denoted as $P_n(t) = \langle n | \sigma_{000000} | n \rangle$ whereas for independent bath case with diagonal fluctuation, population for 7 sites model can be designated as $P_n(t) = \langle n | \sigma_{0000000} | n \rangle$. In the Markovian limit we have compared the QSLE results for same bath and independent bath (off-diagonal fluctuation) case with the results obtained from simple rate equation theory. The rate constants are related to the parameters $V_{od}$ and $b_{od}$ as follows[74-76] (In Markovian limit and for low value of $J$ effect of $J$ on rate constant is negligible),

$$k = \frac{2V_{od}{}^2}{b_{od}}. \tag{20}$$

Please refer to Eq. (38) onwards for the complete derivation of Eq. (20).

1. **Population time correlation function for the independent bath case (off-diagonal fluctuation) and the same bath case**

   A. **Non-Markovian limit**

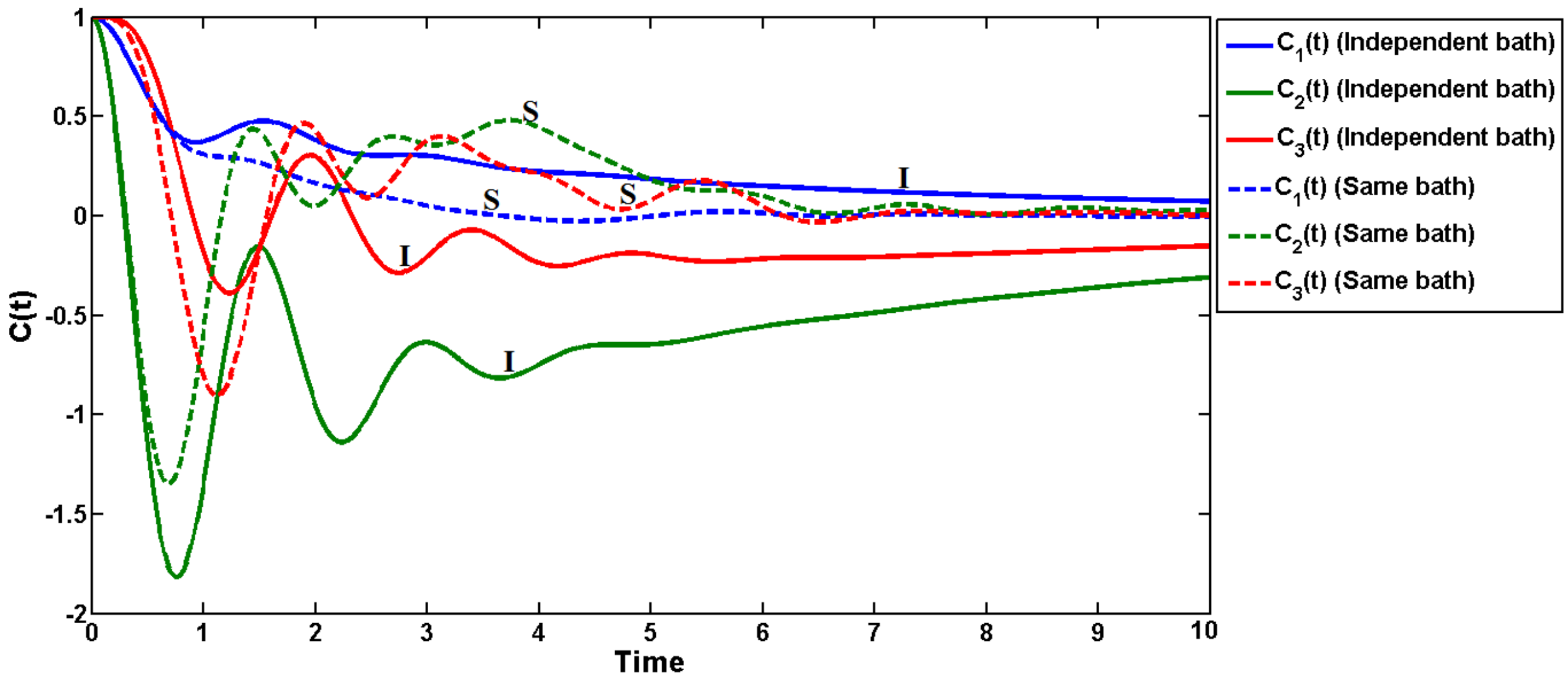

**FIG. 1. Comparison plot of population time correlation functions between the same and the independent bath with off-diagonal fluctuation at $J = V_{od} = b_{od} = 1$ for linear chain of 7 sites model where solid line corresponds to same bath case and dashed line indicates independent bath case with off-diagonal fluctuation. $C_1(t)$, $C_2(t)$ and $C_3(t)$ are the population time correlation function for site 1, site 2 and site 3 respectively. 'I' and 'S' designates independent bath and same results respectively. Oscillatory behavior is superior for same bath case than that of independent bath case with off-diagonal fluctuation as independent bath consists of large number of uncorrelated baths which effectively destroy coherence.**

We have compared PTCF between the same and the independent bath case with off-diagonal fluctuation in non-Markovian limit in Fig. 1 for linear 7 sites model at $J = V_{od} = b_{od} = 1$. The oscillatory behavior of PTCF indicates quantum coherent evolution of superpositions of electronic states. Not only superpositions of electronic states are responsible for the quantum beating but also environment can participate in protecting coherence. Though phase relation between the states is destroyed, environment still has memory of previous state and it reintroduces the phase relation into the system. From the above figures it is clear that the oscillatory behavior persists upto long time for same bath case rather than independent bath case (off-diagonal fluctuation) as large numbers of uncorrelated baths effectively destroy the phase relation between the states for the independent bath case (off-diagonal fluctuation).

### B. Markovian limit

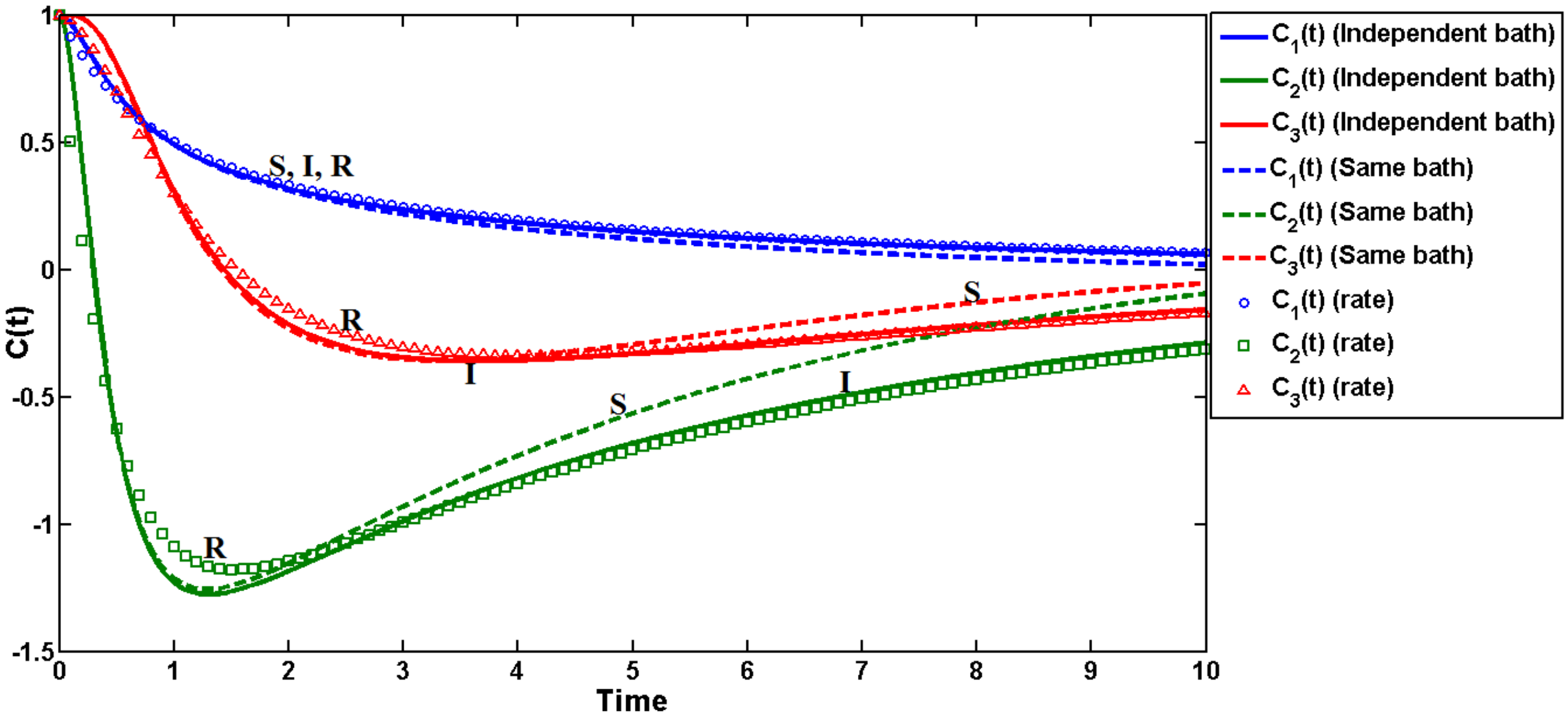

**FIG. 2. Comparison of population time correlation functions between the same bath and the independent bath (off-diagonal fluctuation) with rate theory results for linear chain of 7 sites model is plotted where we have used $J = 0.2, V_{od} = 2$ and $b_{od} = 10$ for same and independent bath case (off-diagonal fluctuation) and rate constant k = 0.8. Solid line corresponds to same bath case, dashed line indicates independent bath case with off-diagonal fluctuation and symbol represents rate theory result. $C_1(t)$, $C_2(t)$ and $C_3(t)$ are the population time correlation function for site 1,**

**site 2 and site 3 respectively. 'I', 'S' and 'R' indicates independent bath, same bath and rate theory results respectively. Agreement between the independent bath (off-diagonal fluctuation) and rate theory result is better than the same bath and rate theory result.**

We have compared PTCF between the same and the independent bath case (only off-diagonal fluctuation) with the PTCF obtained from simple rate equation theory in Markovian limit ( $J = 0.2,\ V_{od} = 2,\ b_{od} = 10$ and $k = 0.8$) in Fig. 2 for linear 7 sites model. In this case it is impossible to reach fully Markovian limit from QSLE description because to get that limit one has to take $V_{od} \to \infty$ and $b_{od} \to \infty$. Though the above parameter space does not fully provide Markovian limit but is good enough to show an excellent agreement between these two approaches. In this limit energy transfer occurs with the equilibrium phonon states and classical rate expression is applicable to explain the exciton energy transfer dynamics. The non-oscillatory decay of PTCF essentially indicates incoherent energy transfer or hopping of the exciton. The agreement between the independent bath case (off-diagonal fluctuation) and rate theory is better than the agreement between the same bath case and rate theory as phase relation still survives between the states in this limit for the same bath case.

1. **Population time correlation function for the independent bath case (diagonal fluctuation)**

**A. Non-Markovian limit**

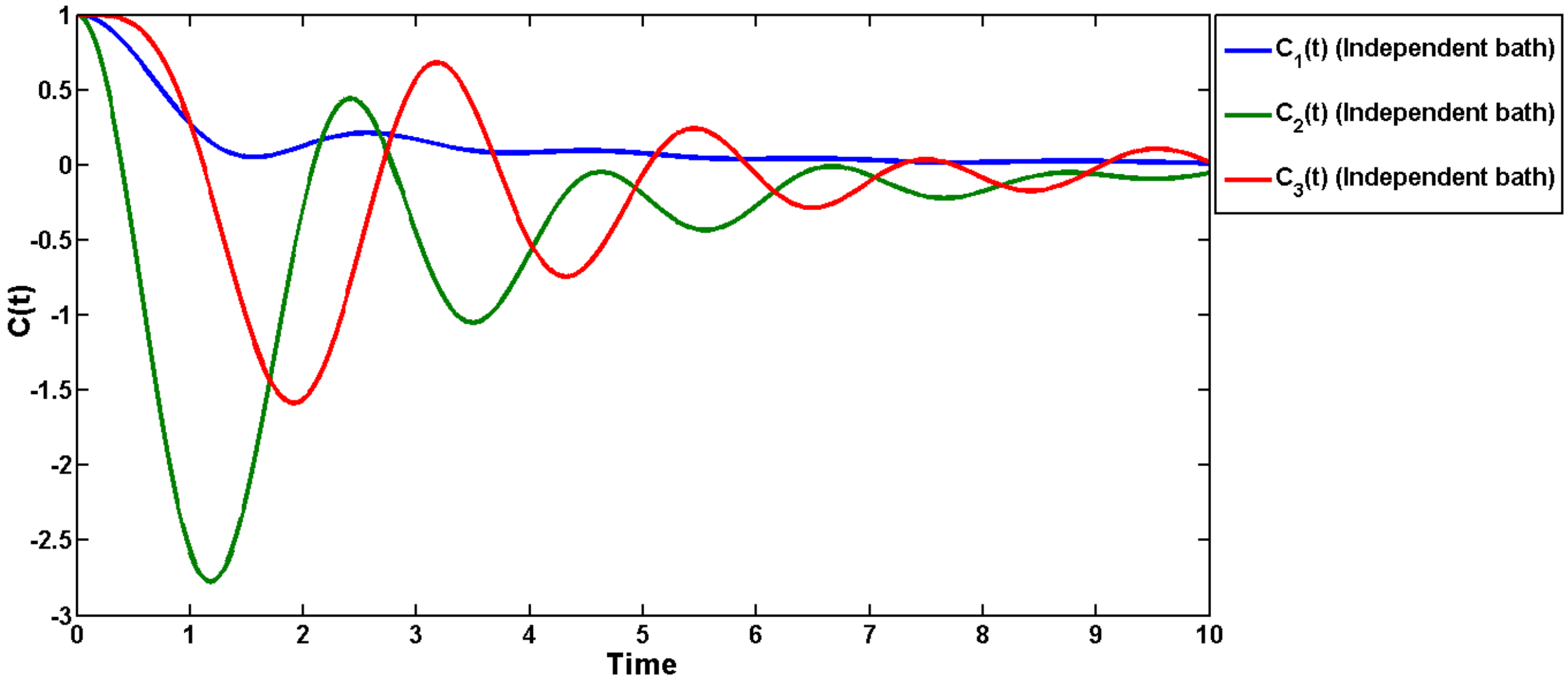

**FIG. 3. Population time correlation functions for the independent bath case with diagonal fluctuation at $J = V_d = b_d = 1$ for linear chain of 7 sites model is plotted against time. $C_1(t)$, $C_2(t)$ and $C_3(t)$ are the population time correlation functions for site 1, site 2 and site 3 respectively. In this case (non-Markovian limit) oscillation is more pronounced than that of independent bath (off-diagonal fluctuation) and same bath case as diagonal fluctuation can't destroy coherence effectively for independent bath (diagonal fluctuation) case.**

We have plotted PTCF for the independent bath case (diagonal fluctuation) at $J = V_d = b_d = 1$ for linear 7 sites model in Fig. 3. For the independent bath case with diagonal fluctuation oscillation is more pronounced than that of same bath and independent bath case with off-diagonal fluctuation in non-Markovian limit. In this case diagonal fluctuation helps to fluctuate site energies whereas off-diagonal fluctuation helps to fluctuate inter-site coupling ( $J$ ).

$J$ is responsible for the oscillation in PTCF and coherent motion of exciton. As coherence can't be directly destroyed by the diagonal fluctuation, coherent transport is most facile in this case.

### B. Markovian limit

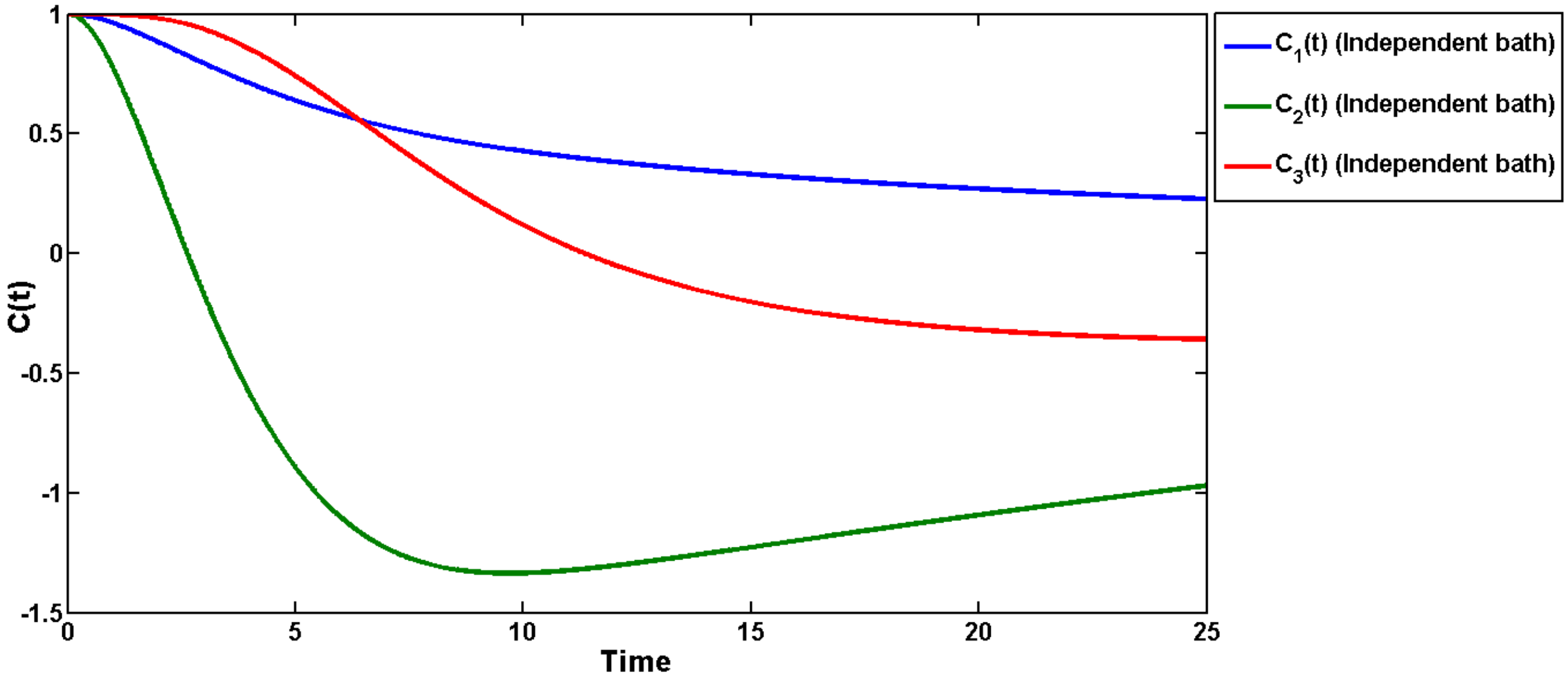

**FIG. 4. Plot shows population time correlation functions for the independent bath case with diagonal fluctuation at $J = 0.2, V_d = 2$ and $b_d = 10$ for linear chain of 7 sites model. $C_1(t)$, $C_2(t)$ and $C_3(t)$ are the population time correlation function for site 1, site 2 and site 3 respectively. Over-damped oscillation is observed in population time correlation function in Markovian limit. Each of the sites requires long time to attain equilibrium population which is due to nature of the bath.**

We have plotted PTCF for the independent bath case (diagonal fluctuation) at $J = 0.2$, $V_d = 2$ and $b_d = 10$ for linear 7 sites model in Fig. 4. For the independent bath case with diagonal fluctuation, we have obtained over-damped PTCF in Markovian limit. In this case all the sites require more time to attain equilibrium population than that of same bath and independent bath case with off-diagonal fluctuation. As coherence can't be directly destroyed by diagonal fluctuation, energy goes back and forth and consequently each site requires long time to obtain equilibrium population.

## V. MEAN SQUARED DISPLACEMENT

Mean squared displacement (MSD) is a tool to obtain the diffusion coefficient of the system. In this work we have calculated MSD for three cases (1) the independent bath with off-diagonal fluctuation (2) the same bath case (3) the independent bath with diagonal fluctuation for discrete model (9 sites model). For all the cases we have considered

Bagchi and Oxtoby calculated MSD for the same bath case for continuum model and used Laplace-Fourier transform method to solve coupled partial differential equations. The expression of MSD is provided as follows

$$\left\langle x^2(t)\right\rangle = 2a^2J^2t^2 + \frac{4a^2{V_{od}}^2}{b_{od}}t + \frac{4a^2{V_{od}}^2}{b_{od}^2}\exp\left(-b_{od}t\right) - \frac{4a^2{V_{od}}^2}{b_{od}^2}\,. \tag{21}$$

For the independent bath case with off-diagonal and diagonal fluctuation separately for long time limit and in the Markovian limit MSD can be expressed as

$$\left\langle x^2(t)\right\rangle \sim 2a^2\left[\left(\frac{2V_{od}^2}{b_{od}}\right) + \frac{J^2}{\left(2V_{od}^2/b_{od}\right)}\right]t \qquad \text{(Considering only off-diagonal fluctuation)} \tag{22}$$

$$\left\langle x^2(t)\right\rangle \sim 2a^2\left[\frac{J^2}{\left(V_d^2/b_d\right)}\right]t\,. \qquad \text{(Considering only diagonal fluctuation)} \tag{23}$$

If one neglects diagonal fluctuation the Haken-Reineker expression of diffusion coefficient can be written as follows

$$D_{HR} = a^2 \left[ 2\gamma_{od} + \frac{J^2}{2\gamma_{od}} \right]. \qquad (24)$$

If one considers only diagonal fluctuation then diffusion coefficient can be expressed as

$$D_{HR} = a^2 \frac{J^2}{\gamma_d}. \qquad (25)$$

Eq. (22) and Eq. (24); Eq. (23) and Eq. (25) are same. We have already explained the relation between $\gamma_d$, $V_d$ and $b_d$ as well as the relation between $\gamma_{od}$, $V_{od}$ and $b_{od}$. We have compared diffusion coefficient of our discrete model with Haken-Reineker diffusion coefficient for continuum model.

1. **Mean squared displacement for the independent bath case (off-diagonal fluctuation) and the same bath case**

   **A. Non-Markovian limit**

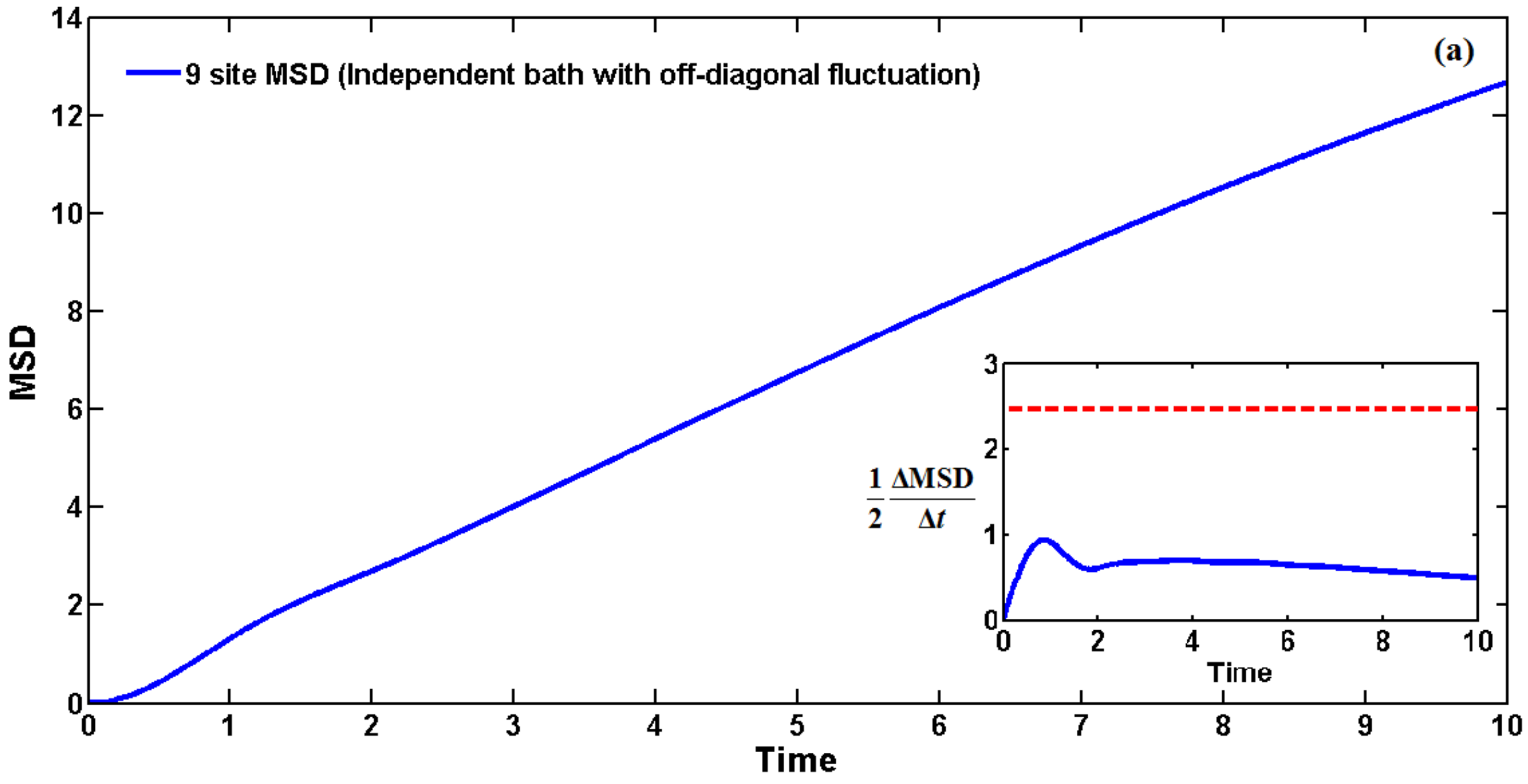

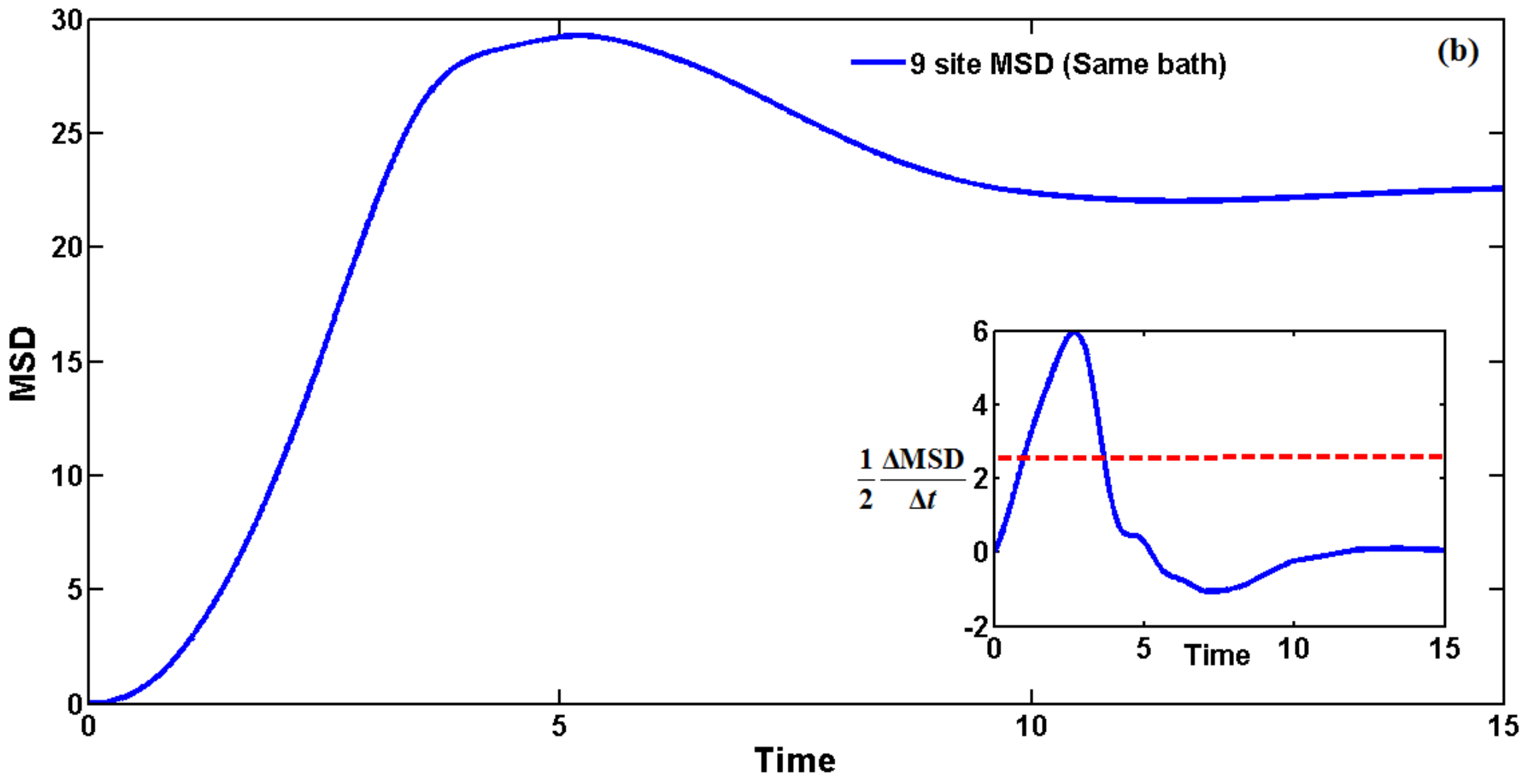

**FIG. 5 Mean squared displacement is plotted for linear chain of 9 sites model for the independent bath (off-diagonal fluctuation) and the same bath case in non-Markovian limit for (a) independent bath case with off-diagonal fluctuation at $J = V_{od} = b_{od} = 1$ and (b) same bath case at $J = V_{od} = b_{od} = 1$. For both cases half of the derivative plot of mean squared displacement is provided as inset where red dashed line indicates Haken-Reineker diffusion coefficient. No flat region in the half of the derivative plot of mean squared displacement indicates coherent transport of exciton for both same and independent bath (off-diagonal fluctuation) case.**

Fig. 5(a) and Fig. 5(b) signify MSD plots for the independent bath (off-diagonal fluctuation) and the same bath case respectively at $J = V_{od} = b_{od} = 1$ for linear 9 sites model. Half of the derivative of MSD plot is provided as inset for both the figures. In the non-Markovian limit no plateau in half of the derivative of MSD plots in case of both independent bath with off-diagonal fluctuation and same bath clearly indicates coherent transport for all the cases. Transport is more coherent for the same bath case than that of the independent bath case (off-diagonal fluctuation) due to the presence of large number of uncorrelated baths in case of independent bath. From PTCF plot it is also evident that oscillatory dynamics of the exciton correspond to the coherent transport. From PTCF plot we have also observed that oscillation is superior for same bath case than that of independent bath case (off-diagonal fluctuation).

## B. Markovian limit

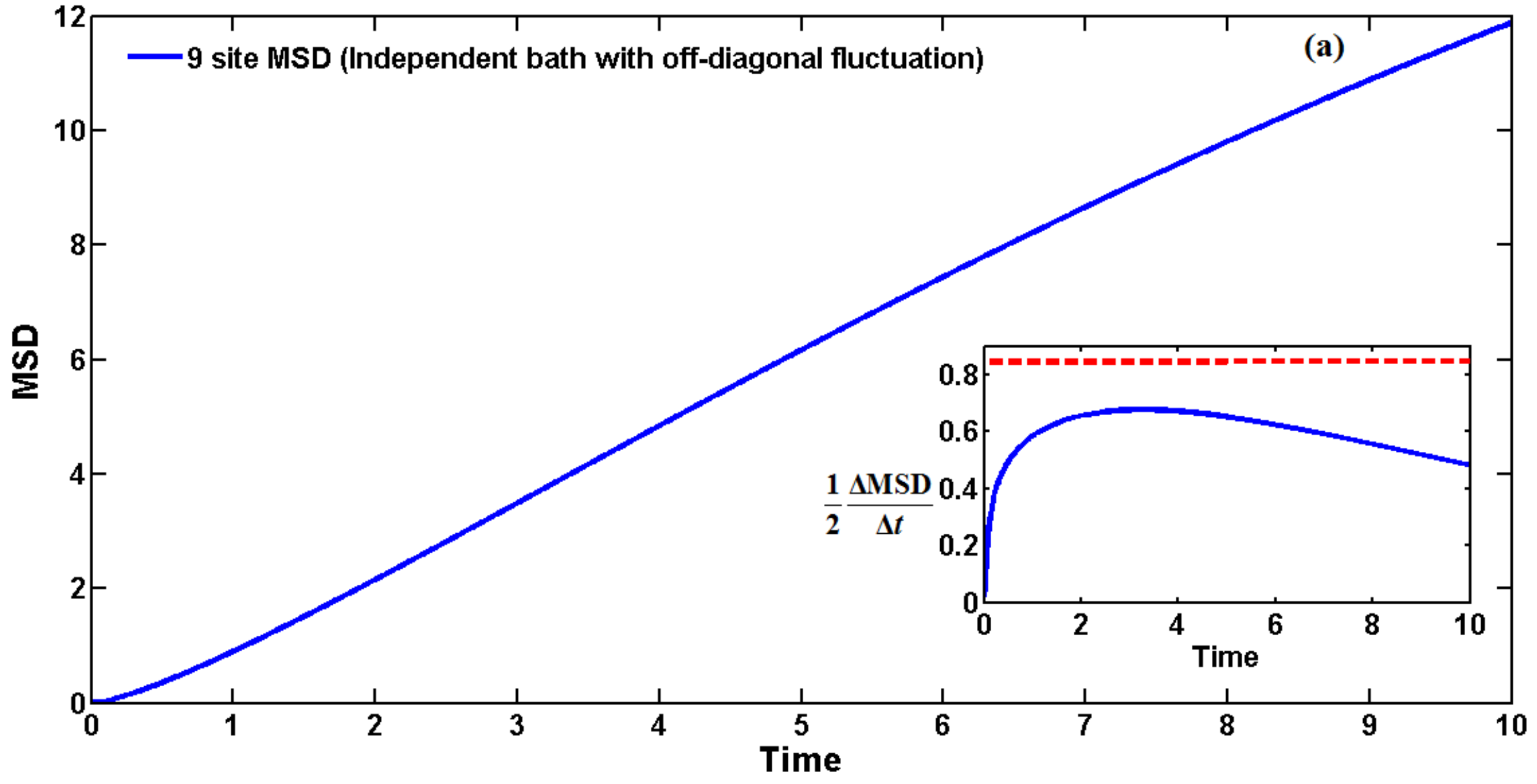

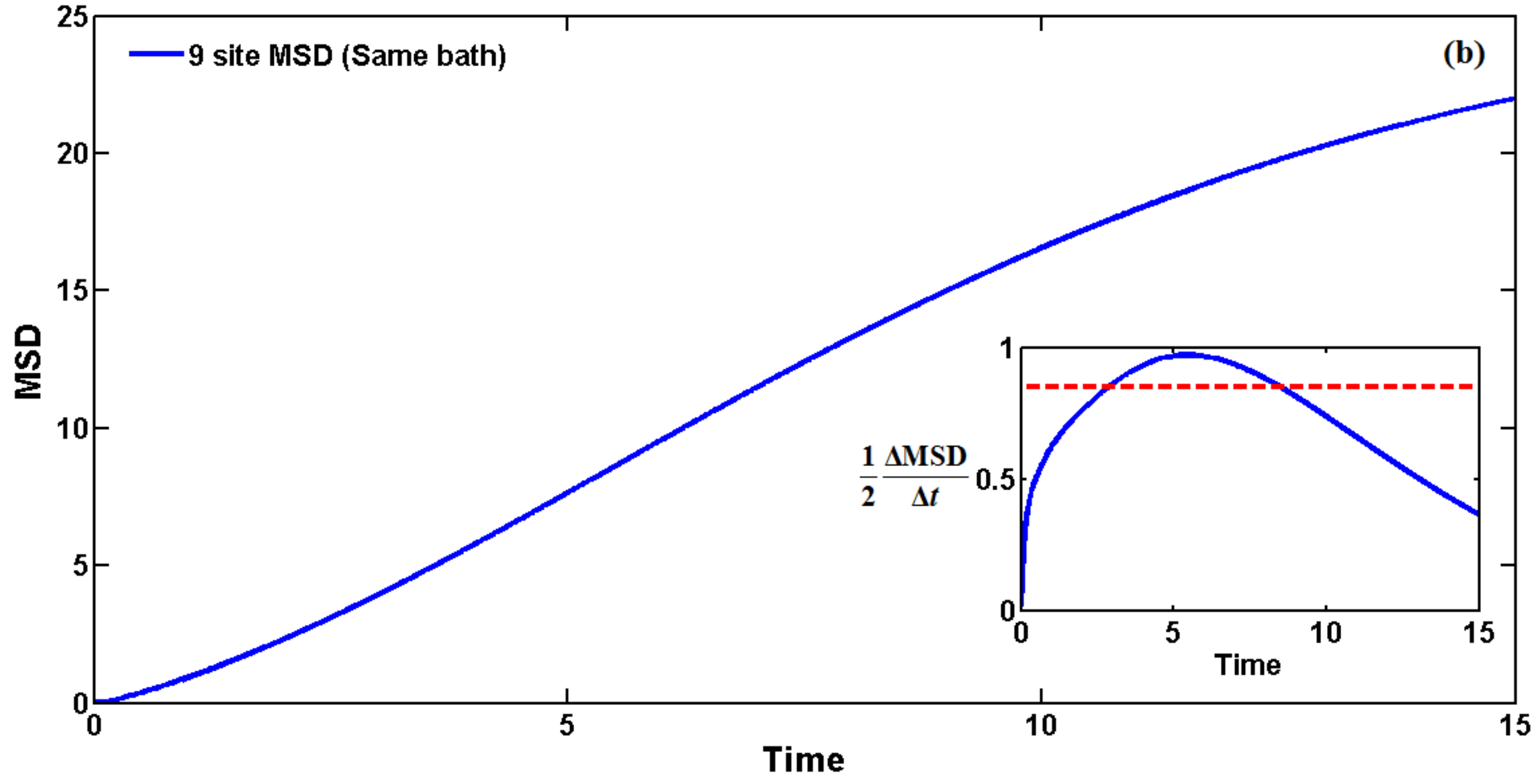

**FIG. 6 Mean squared displacement is plotted for linear chain of 9 sites model for the independent bath (off-diagonal fluctuation) and same bath case in Markovian limit for (a) independent bath case**

**with off-diagonal fluctuation at $J = 0.2, V_{od} = 2$ and $b_{od} = 10$ as well as (b) same bath case at $J = 0.2, V_{od} = 2$ and $b_{od} = 10$. For both the figures half of the derivative plot of mean squared displacement is provided as inset where red dashed line indicates Haken-Reineker diffusion coefficient. The diffusive behavior or incoherent transport is clear from linear behavior of mean squared displacement and plateau in the half of the derivative plot of mean squared displacement.**

Fig. 6(a) and Fig. 6(b) show the plots of MSD for independent bath (off-diagonal fluctuation) and same bath case respectively at $J = 0.2, V_{od} = 2$ and $b_{od} = 10$ for linear 9 sites model. Half of the derivative of MSD plot is provided as inset for both the figures. For the independent bath case with off-diagonal fluctuation in Markovian limit, we have obtained flat region in the half of the derivative of MSD plot of 9 sites model which suggest diffusive motion of exciton. The MSD plot consists of three regions in the Markovian limit. Near $t = 0$ low value of half of the derivative of MSD indicates ballistic or coherent motion of exciton. After that linear region in MSD corresponds exciton diffusion followed by the saturation of MSD. For independent bath case (off-diagonal) with increasing system size one can obtain linear behavior of MSD which survives upto long time. Difference between MSD of independent bath (off-diagonal fluctuation) and the same bath is less in this limit (for $J = 0$ with high $V_{od}$ and $b_{od}$ one will obtain same MSD). However, this difference in the Markovian limit will increase with increasing system size. We have observed from PTCF plot that in this limit the difference between the independent bath case (off-diagonal fluctuation) and the same bath case is also less. It is interesting to note that the independent bath (off-diagonal fluctuation) case shows diffusive motion of exciton in this limit whereas for same bath case no signature of diffusion is observed.

## 2. Mean squared displacement for the independent bath case (diagonal fluctuation)

### A. Non-Markovian limit

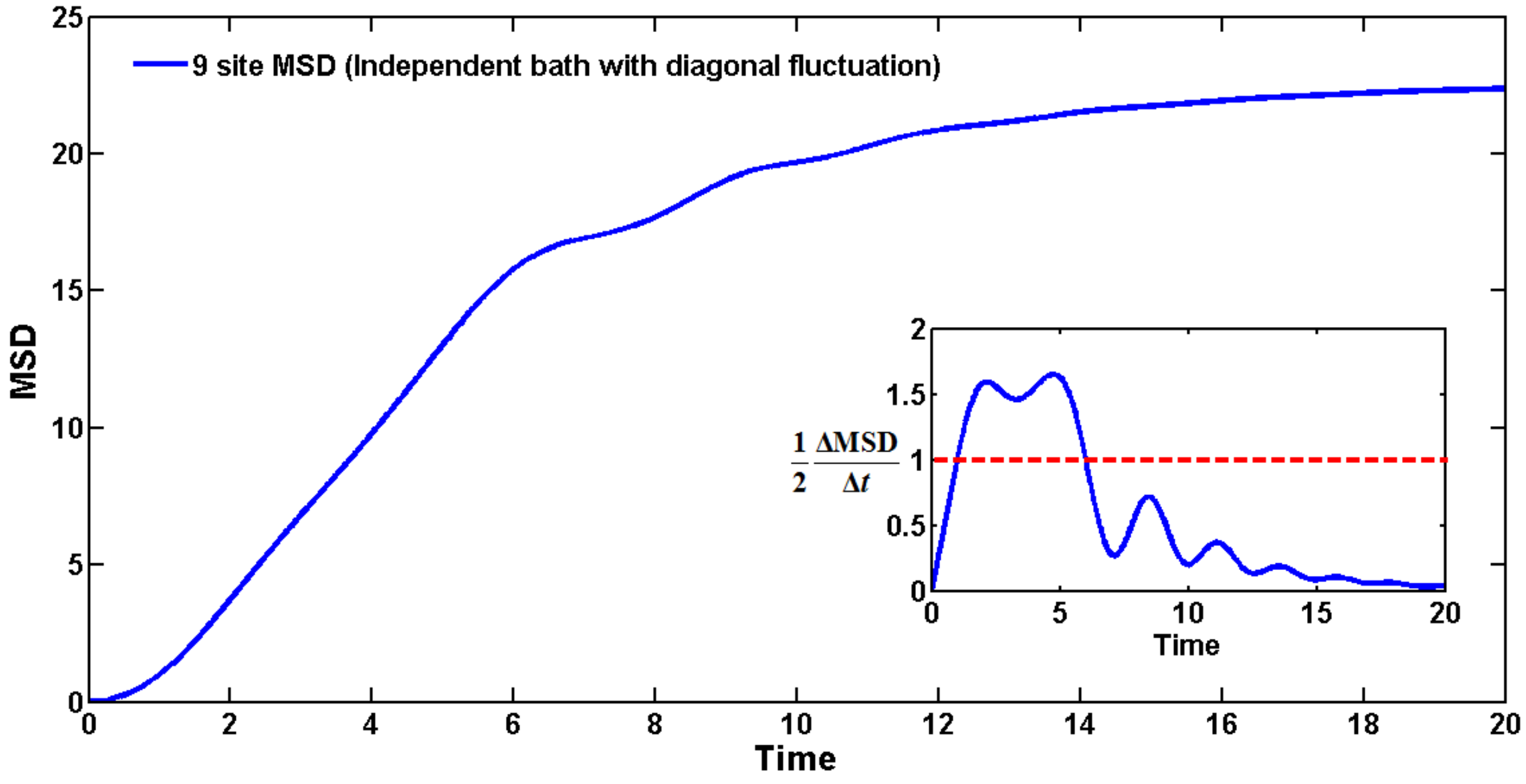

**FIG. 7. Mean squared displacement is plotted for the independent bath case with diagonal fluctuation at $J = V_d = b_d = 1$ for linear chain of 9 sites model. Half of the derivative plot of mean squared displacement is provided as inset where red dashed line indicates Haken-Reineker diffusion coefficient. No flat region in the half of the derivative plot suggests coherent transport of exciton. In this case transport is more coherent than the same bath case and independent bath case with off-diagonal fluctuation.**

In Fig. 7 we have plotted MSD for independent bath (diagonal fluctuation) at $J = V_d = b_d = 1$ for 9 sites linear model and we also have plotted half of the derivative plot of MSD as inset. Absence of flat region in half of the derivative plot of MSD in case of independent bath with diagonal fluctuation indicates fully coherent transport in non-Markovian limit. Diagonal fluctuation helps to fluctuate site energy and can't effectively destroy coherence.

Consequently in this case coherent transport is easiest. We have also understood from the earlier plot of PTCF that in non-Markovian limit oscillation is more pronounced in case of independent bath with diagonal fluctuation than that of the independent bath case with off-diagonal fluctuation and same bath case.

## B. Markovian limit

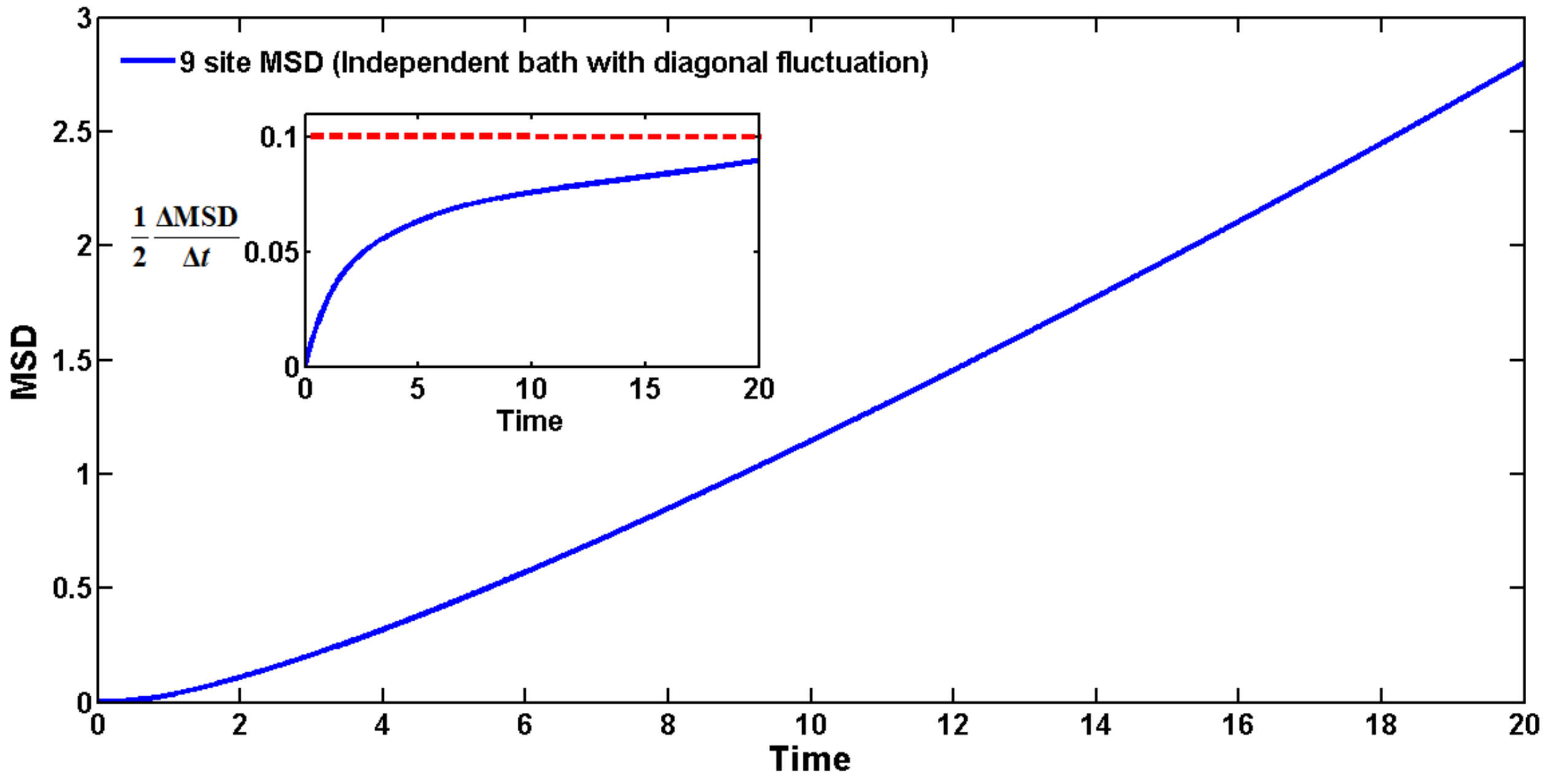

**FIG. 8. Mean squared displacement for the independent bath case with diagonal fluctuation at $J = 0.2, V_d = 2$ and $b_d = 10$ for linear chain of 9 sites model. Half of the derivative plot of mean squared displacement is provided as inset where red dashed line indicates Haken-Reineker diffusion coefficient. Half of the derivative plot of mean squared displacement shows after initial ballistic motion of exciton diffusion occurs and it stays upto long time. The diffusive behavior or incoherent transport is clear from linear behavior of mean squared displacement and plateau in the half of the derivative plot of mean squared displacement.**

In Fig. 8 we have represented MSD plot for independent bath (diagonal fluctuation) at $J = 0.2,\ V_d = 2$ and $b_d = 10$ for 9 sites linear model and we have provided half of the derivative plot of MSD as inset. For the independent bath with diagonal fluctuation only energy of each site can fluctuate randomly due to the diagonal fluctuation. Long time is required to abolish the coherence. That is why coherent transport in initial regime continues upto long time. We have obtained a linear region in MSD plot which corresponds to diffusive or incoherent motion of exciton. From the earlier PTCF plot we also have observed over-damped oscillation in Markovian limit and each of the sites requires long time to attain equilibrium population.

## VI. RELATIONSHIP BETWEEN COHERENCE AND QUANTUM DIFFUSION

In classical systems, a particle coupled to a stochastic bath with dissipation usually exhibits diffusive behavior in the long time, even in the non-Markovian limit. Notable exceptions appear when the motion cannot be reduced to a random walk, as in Levy flight where the second moment of the time dependent population distribution does not exist, and the mean squared displacement is super-diffusive. In the other extreme of motion on a fractal object, mean squared displacement can be sub-diffusive.

In the system studied here, where the particle or excitation moves on a regular lattice, diffusion is expected in the long time. In the classical limit, a particle can hop from site to site when there is a non-zero coupling between different sites so that a rate of transfer exists. This situation changes completely for quantum particles where coherence can prevent diffusive behavior to set in, even in a long time.

The presence of coherence is reflected in the time dependence of population which exhibits an oscillatory back and forth behavior, as shown in **Figs.** 1 and 3 of this work. The oscillatory decay of population essentially indicates transfer of coherence. A theoretical analysis of the same bath case provides fruitful insight into the propagation of coherence. In this case, the coupled equation of motion is given by the following set of coupled equations

$$\begin{aligned}\langle 2|\dot{\sigma}_0|2\rangle &= -iJ\langle 1|\sigma_0|2\rangle + iJ\langle 2|\sigma_0|1\rangle - iJ\langle 3|\sigma_0|2\rangle + iJ\langle 2|\sigma_0|3\rangle - iV_{od}\langle 1|\sigma_1|2\rangle \\ &\quad + iV_{od}\langle 2|\sigma_1|1\rangle - iV_{od}\langle 3|\sigma_1|2\rangle + iV_{od}\langle 2|\sigma_1|3\rangle \end{aligned} \tag{26a}$$

$$\begin{aligned}\langle 2|\dot{\sigma}_1|2\rangle &= -iJ\langle 1|\sigma_1|2\rangle + iJ\langle 2|\sigma_1|1\rangle - iJ\langle 3|\sigma_1|2\rangle + iJ\langle 2|\sigma_1|3\rangle - iV_{od}\langle 1|\sigma_0|2\rangle \\ &\quad + iV_{od}\langle 2|\sigma_0|1\rangle - iV_{od}\langle 3|\sigma_0|2\rangle + iV_{od}\langle 2|\sigma_0|3\rangle - b_{od}\langle 2|\sigma_1|2\rangle \end{aligned} \tag{26b}$$

$$\langle 1|\dot{\sigma}_0|2\rangle = -iJ\langle 2|\sigma_0|2\rangle + iJ\langle 1|\sigma_0|1\rangle - iV_{od}\langle 2|\sigma_1|2\rangle + iV_{od}\langle 1|\sigma_1|1\rangle \tag{26c}$$

$$\langle 1|\dot{\sigma}_1|2\rangle = -iJ\langle 2|\sigma_1|2\rangle + iJ\langle 1|\sigma_1|1\rangle - iV_{od}\langle 2|\sigma_0|2\rangle + iV_{od}\langle 1|\sigma_0|1\rangle - b_{od}\langle 1|\sigma_1|2\rangle \tag{26d}$$

$$\langle 2|\dot{\sigma}_0|3\rangle = iJ\langle 2|\sigma_0|2\rangle - iJ\langle 3|\sigma_0|3\rangle + iV_{od}\langle 2|\sigma_1|2\rangle - iV_{od}\langle 3|\sigma_1|3\rangle \tag{26e}$$

$$\langle 2|\dot{\sigma}_1|3\rangle = iJ\langle 2|\sigma_1|2\rangle - iJ\langle 3|\sigma_1|3\rangle + iV_{od}\langle 2|\sigma_0|2\rangle - iV_{od}\langle 3|\sigma_0|3\rangle - b_{od}\langle 2|\sigma_1|3\rangle \,. \tag{26f}$$

As shown in **Ref.** (29), these coupled equations can be reduced in the continuum limit to the following set of partial differential equations (where the dots represent time derivatives while the primes denote spatial derivatives)

$$\ddot{P}_0 + M\dot{P}_0 = AP_0'' + BP_1'' - EZ_0 \tag{27a}$$

$$\ddot{P}_1 + M\dot{P}_1 = AP_1'' + BP_0'' + EZ_1 \tag{27b}$$

$$\dot{Z}_0 = CP_0'' + DP_1'' \tag{27c}$$

$$\dot{Z}_1 = CP_1'' + DP_0'' - MZ_1 \tag{27d}$$

where $$A = 2a^2\left(J^2 + V_{od}{}^2\right) \tag{28a}$$

$$B = 4JV_{od}a^2 \tag{28b}$$

$$C = 2iJa^2 \tag{28c}$$

$$D = 2iV_{od}a^2 \tag{28d}$$

$$E = iJb_{od} \tag{28e}$$

$$M = b_{od} \tag{28f}$$

and $a$ is the lattice spacing. The other quantities are defined as follows

$$P_0(x,t) = \langle x|\sigma_0(t)|x\rangle \tag{29a}$$

$$P_1(x,t) = \langle x|\sigma_1(t)|x\rangle \tag{29b}$$

$$Z_0(x,t) = \langle x+a|\sigma_0(t)|x\rangle + \langle x-a|\sigma_0(t)|x\rangle - \langle x|\sigma_0(t)|x+a\rangle - \langle x|\sigma_0(t)|x-a\rangle \tag{29c}$$

$$Z_1(x,t) = \langle x+a|\sigma_1(t)|x\rangle + \langle x-a|\sigma_1(t)|x\rangle - \langle x|\sigma_1(t)|x+a\rangle - \langle x|\sigma_1(t)|x-a\rangle \tag{29d}$$

In continuum limit $P_n(t)$ can be replaced by $P(x,t)$ where $x = na$.

Eq. (27) is a coupled partial differential equation and is highly complex. Only in the limit $J = 0$ for the same bath case, the exciton motion is diffusive. This can be understood by analyzing equations Eq. (26) and Eq. (27) as follows. If the exciton density does not change rapidly one can neglect the second derivative from left hand side in Eq. (25a). Since $B$ and $E$ both are zero when $J=0$, we recover diffusion of the exciton with the diffusion coefficient $\frac{2V_{od}^2}{b_{od}}$. This is precisely same as the Haken-Reineker diffusion coefficient at $J=0$.

Even from Eq. (26) and Eq. (27) one can easily conclude that coherence is propagated through the excited bath states i.e. $\sigma_1$. In non-Markovian limit, oscillatory decay of population comes from coherence propagated through the excited bath state where transport is coherent.

For system to become incoherent, the excited bath state population should decay fast. In Eq. (27), we recover a diffusion equation easily when (i) $J=0$, (ii) $V_{od}$ and $b_{od}$ is large and at long time limit.

For the independent bath case we have taken $J=0$ and also considered only off-diagonal fluctuation. Hence from **Ref.** (29) one can write coupled equation of motion for continuum model as follows

$$\dot{P}_0 = -iV_{od}Z_1 \tag{30a}$$

$$\dot{Z}_1 = 2iV_{od}a^2P_0'' - b_{od}Z_1 \tag{30b}$$

where $P_0(x,t) = \langle x|\sigma_{00000}(t)|x\rangle$ (31a)

$$Z_1(x,t) = \langle x+a|\sigma_{00001}(t)|x\rangle + \langle x-a|\sigma_{00010}(t)|x\rangle - \langle x|\sigma_{00001}(t)|x+a\rangle - \langle x|\sigma_{00010}(t)|x-a\rangle . \tag{31b}$$

From the Laplace transformation of Eq. (30a) and Eq. (30b) we obtain

$$s\tilde{P}_0\left(x,s\right)-\delta(x)=-iV_{od}a^2\tilde{Z}_1(x,s) \tag{32a}$$

$$\tilde{Z}_1(x,s)=\frac{2iV_{od}a^2}{\left(s+b_{od}\right)}\tilde{P}_0''(x,s)\,. \tag{32b}$$

Coherence is propagated through excited bath state i.e. the term $Z_1$ is responsible for coherence. If $Z_1=0$ we get back the diffusion equation.

Substituting Eq. (32b) into Eq. (33a) we obtain

$$s\tilde{P}_0\left(x,s\right)-\delta(x)=\frac{2V_{od}a^2}{\left(s+b_{od}\right)}\tilde{P}_0''\left(x,s\right). \tag{33}$$

From the Fourier transformation of Eq. (33) followed by rearrangement of same we get

$$\tilde{P}_0\left(k,s\right)=\frac{\left(s+b_{od}\right)}{s\left(s+b_{od}\right)+2V_{od}a^2k^2}\,. \tag{34}$$

We know that,

$$\left\langle x^2(s)\right\rangle=-\left.\frac{\partial^2\tilde{P}_0(k,s)}{\partial k^2}\right|_{k=0}\,. \tag{35}$$

We use Eq. (34) and Eq. (35) and eventually inverse Laplace transform of the resulting equation yields,

$$\left\langle x^2(t)\right\rangle=\frac{4a^2V_{od}{}^2}{b_{od}}t+\frac{4a^2V_{od}{}^2}{b_{od}^2}\exp\left(-b_{od}t\right)-\frac{4a^2V_{od}{}^2}{b_{od}^2}\,. \tag{36}$$

We find diffusive behavior to emerge in the time going to the infinity limit as the first term dominates in this limit. This gives rise to a value of the diffusion coefficient as $\frac{2a^2 V_{od}{}^2}{b_{od}}$.

Eq. (36) is same as Eq. (21) for the case of $J = 0$ which suggests that for both the same and the independent bath case (off-diagonal fluctuation) MSD behaves in a similar manner. In this limit we have also obtained similar result from our numerical solution of the discrete models. We have recovered the diffusive behavior of exciton for both the same bath (long chain) and the independent bath case at long time limit for $J = 0$ from continuum model and also from discrete model before the saturation of MSD.

## VII. DIFFUSION OF EXCITON IN LONG CHAIN LIMIT

Study of exciton transfer in a long chain (large number of site) is complicated because numerical calculation is computationally expensive. However it becomes simpler for the same bath case when $J = 0$. Same bath case is computationally less expensive than that of the independent bath case due to the less number of coupled differential equation for the same bath than that of the independent bath. We have considered same bath case upto 51 sites and independent bath case upto 9 sites. The main motivation for considering large system is to obtain diffusion coefficient close to the analytical results for continuum model.

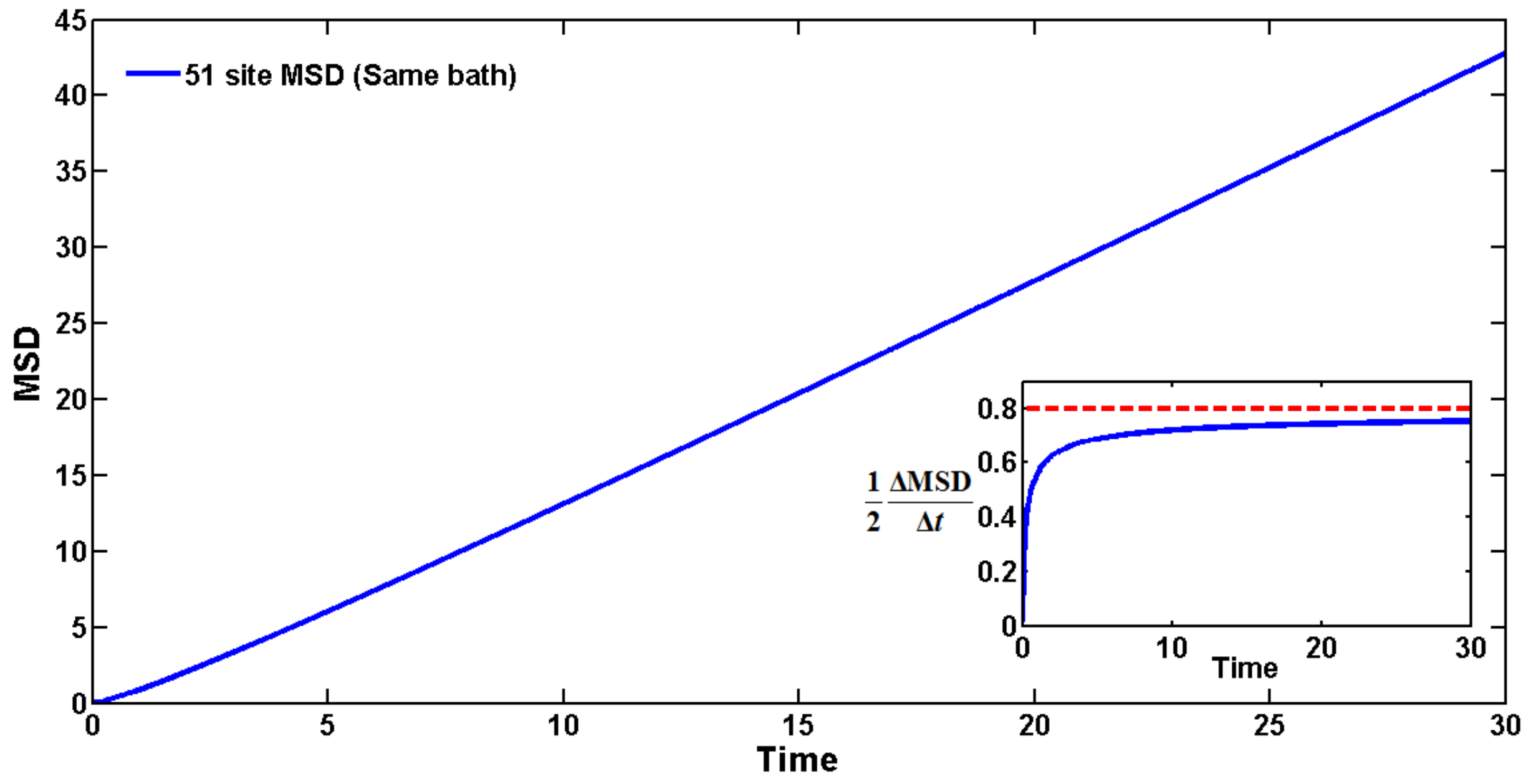

**FIG. 9. Mean squared displacement is plotted for the same bath case at $J = 0, V_{od} = 2$ and $b_{od} = 10$ for linear chain of 51 sites model. Half of the derivative plot of mean squared displacement is provided as inset where red dashed line indicates Haken-Reineker diffusion coefficient. The diffusive behavior or incoherent transport is clear from linear behavior of mean squared displacement and plateau in the half of the derivative plot of mean squared displacement. Diffusion coefficient obtained from discrete model calculation shows good agreement with the Haken-Reineker diffusion coefficient for continuum model.**

We have plotted MSD for the same bath case at $J = 0,\ V_{od} = 2$ and $b_{od} = 10$ for 51 sites linear model in Fig. 9 and we have provided half of the derivative plot of MSD as inset. From Fig. 9 we can conclude that with increasing system size (large number of sites) the diffusion coefficient obtained from the discrete model system will be close to the diffusion coefficient obtained from analytical calculation for the continuum model. Also another interesting feature of Fig. 5 is that the linear region in MSD plot continues upto long time before the saturation of MSD which is also clear from long plateau region in half of the derivative of MSD plot.

## VIII. CALCULATION OF RATE OF POPULATION TRANSFER AND DIFFUSION COEFFICIENT

One can carry out an analysis that makes connection with classical limit of quantum diffusion. For one dimensional random walk[2], diffusion coefficient can be expressed as follows

$$D_{RW} = \frac{1}{2} k a^2 \tag{37}$$

where $k$ is the rate of hopping and $a$ is the length of each step. We have used the derivation of Oxtoby[4,74-76] for the calculation of rate of hopping.

The rate for the transfer process from $i^{th}$ states to the $j^{th}$ one, can be expressed by the use of Fermi-Golden rule as follows,

$$k_{ij} = \frac{2\pi}{\hbar^2} \sum_{\alpha} \sum_{\beta} P_{\alpha} \left| V_{i\alpha, j\beta} \right|^2 \delta\left( E_i - E_j + E_{\alpha} - E_{\beta} \right). \tag{38}$$

where $i$ and $j$ are the eigen-states of the system; $\alpha$ and $\beta$ are the eigen-states of the bath.

Using the Fourier transform of delta function, Eq. (38) can be written as,

$$k_{ij} = \frac{1}{\hbar^2} \int_{-\infty}^{\infty} dt \exp\left( i\omega_{ij} t \right) \sum_{\alpha} \sum_{\beta} P_{\alpha} \exp\left( \frac{it}{\hbar} E_{\alpha} \right) V_{i\alpha, j\beta} \otimes \exp\left( -\frac{it}{\hbar} E_{\beta} \right) V_{j\beta, i\alpha} \tag{39}$$

One can transform Eq. (39) employing Heisenberg representation as follows,

$$k_{ij} = \int_{-\infty}^{\infty} dt \exp\left( i\omega_{ij} t \right) \left\langle V_{ij}\left( t \right) V_{ji}\left( 0 \right) \right\rangle \tag{40}$$

Eq. (40) can be written in the time symmetrized anti-commutator as,

$$k_{ij} = \frac{2\hbar^{-2}}{1+\exp\left(-\beta\hbar\omega_{ij}\right)} \int_{-\infty}^{\infty} dt \exp\left(i\omega_{ij}t\right) \otimes \left\langle \frac{1}{2}\left[V_{ij}\left(t\right), V_{ji}\left(0\right)\right]_{+} \right\rangle \tag{41}$$

where anti-commutaror $\left[A,B\right]_{+} = AB + BA$.

Using semi-classical approximation (by replacing the anti-commutator by classical correlation function) one can obtain the rate constant for the transfer process as follows

$$k_{ij} = \frac{2\hbar^{-2}}{1+\exp\left(-\beta\hbar\omega_{ij}\right)} \int_{-\infty}^{\infty} dt \exp\left(i\omega_{ij}t\right) \left\langle V_{ij}^{class}\left(t\right) V_{ji}^{class}\left(0\right) \right\rangle \tag{42}$$

where $k_{ij}$ is the rate of transfer from state $i$ to state $j$, $\omega_{ij}$ is the frequency difference between the states and $\beta = \left(k_B T\right)^{-1}$. If the stochastic force is either Gaussian or Poisson, the correlation function for Markov process can be written as,

$$\left\langle V_{ij}(t) V_{ji}(0) \right\rangle = \left\langle V_{ij}^{2}(0) \exp\left(-bt\right) \right\rangle = V_{od}^{2} \exp\left(-b_{od} t\right). \tag{43}$$

For high temperature or classical approximation and for two states with same energy, the rate constant for the transfer process can be written as ($\hbar = 1$)

$$k = \frac{2V_{od}^{2}}{b_{od}}. \tag{44}$$

We already have discussed that PTCF obtained from rate equation theory (rate constant is obtained from Eq. (44)) is close to the PTCF obtained from QSLE in the Markovian limit. Substituting Eq. (44) into Eq. (38) we obtain the diffusion coefficient as follows,

$$D_{RW} = \frac{V_{od}^{2}}{b_{od}} a^{2} = \gamma_{od} a^{2} \tag{45}$$

As discussed in **Sec. VI**, in Markovian limit and at $J=0$, the diffusion coefficient for the continuum model can be reduced to the following expression (from Eq. (6)) as,

$$D_{HR}=2\gamma_{od}a^2 \tag{46}$$

## IX. CONCLUSION

In this work we study coherent and incoherent energy transfer in linear chain of discrete model system with both diagonal and off-diagonal disorder. We have calculated population transfer dynamics (by PTCF) in both non-Markovian and Markovian limits by employing Kubo's stochastic Liouville equation (QSLE) which offers a quantitative approach to solve this difficult problem. We also have computed MSD in the above limits. For the same bath limit with off-diagonal disorder, we have considered linear chain with maximum of 51 sites. However, for the independent bath, we could study only upto for 9 site model.

Let us summarize the main results of the work.

1. In the non-Markovian limit, the population transfer time correlation function (PTCF) shows oscillatory behavior which is a signature of coherent energy transfer. The oscillatory behavior survives for a longer time for the same bath case than that of independent bath case (off-diagonal fluctuation). For the independent bath case, large numbers of uncorrelated baths effectively destroy the phase relation between the states. For the independent bath case with diagonal fluctuation oscillation is more prominent than same bath and independent bath case (off-diagonal fluctuation) as diagonal fluctuation can't destroy coherence directly.

2. We have shown that coherence is propagated through the excited bath state in the present formulation. This point deserves further study. Here the excited bath state contains information of the system not only through the site location but also through the averaging performed over the system density matrix, $\rho$, as shown by Eq. (11).
3. In the Markovian limit, non-oscillatory decay of PTCF indicates energy transfer through incoherent hopping mechanism. In this limit quantum superposition is completely destroyed and energy transfer occurs through equilibrium phonon states. In this case one can use classical random walk or rate theories to explain the energy transfer dynamics. We have also evaluated PTCF from simple rate equation to compare the result with QSLE description (both same and independent bath case (off-diagonal fluctuation)) in this limit. Though it is impossible to reach proper Markovian limit from our model, our parameter space $J = 0.2, V_{od} = 2$ and $b_{od} = 10$ provide satisfactory agreement with rate theory result. But same bath case still has some differences with rate theory which are due to the presence of coherence. For independent bath with diagonal fluctuation in Markovian limit we have obtained over-damped PTCF which suggest slow transition from coherent to incoherent motion. In this case each of the sites requires long time to attain equilibrium population.
4. In the non-Markovian limit absence of any flat region in half of the derivative of MSD plots indicates absence of diffusion in the finite sized system studied here and implies coherent transport for all the cases. Transport is more coherent for the same bath case than that of the independent bath case with off-diagonal fluctuation due to the presence of large number of uncorrelated baths in case of independent bath. In the case of independent bath with diagonal fluctuation, transport is fully coherent and oscillation occurs with large amplitude. Because in this case diagonal fluctuation can't destroy the coherence directly.

5. To obtain the Markovian limit and incoherent transport we have to consider two limiting conditions simultaneously (a) $J$ should be less than both $V$ (either $V_d$ or $V_{od}$) and $b$ (either $b_d$ or $b_{od}$) (b) The ratio of $V^2/b$ should be greater than $2J$. For the independent bath case with off-diagonal fluctuation in the Markovian limit we have recovered diffusive behavior. In this case after initial ballistic motion exciton transport is incoherent. In Markovian limit the difference of MSD between same and independent bath (off-diagonal fluctuation) is less. In a short chain system the difference is less but with increasing system size difference will increase. We have also analytically shown (the coupled equations describe all the real lattice correctly except at very short times) that at $J=0$ both same and independent bath (off-diagonal fluctuation) provide same expression which we have also noticed from numerical calculation. For the same bath case at $J=0$ diffusive behavior of exciton in long chain (51 sites) continues upto long time and diffusion coefficient is very close to the Haken-Reniker diffusion coefficient. In Markovian limit MSD of independent bath with diagonal fluctuation shows diffusive behavior in long time limit before the saturation of MSD. Though the limit is Markovian, coherent motion of exciton lasts upto long time before the onset of diffusive or incoherent transport of exciton.

The main insights derived from the preceding study can be articulated as follows (1) In the Markovian limit, same bath and independent bath (off-diagonal fluctuation) provide similar PTCF and also MSD for smaller chains. (2) We have recovered a diffusive behavior for the independent bath (off-diagonal fluctuation) in the Markovian regime but for same bath case transport is coherent. (3) In the non-Markovian limit, both the independent bath (off-diagonal fluctuation) and the same bath results are profoundly different. Exciton transport remains coherent for both the cases, but the characteristics of PTCF and MSD are different.

In this work we have mainly focused on coherent vs. incoherent migration of exciton in non-Markovian and Markovian limit. Furthermore, one of the important aspects in exciton migration is the efficiency of exciton transfer. One can find high efficiency of exciton migration in case of partial coherent and incoherent transport. Cao and Silbey[49] have calculated the efficiency for small systems. In future we will extend our study to determine the efficiency of large system using QSLE approach.

## ACKNOWLEDGMENTS

We thank Dr. Sarmistha Sarkar and Tuhin Samanta for help during preparation of the manuscript. BB thanks Department of Science and Technology (DST, India) and Sir J. C. Bose fellowship for providing partial financial support.

## APPENDIX A: DERIVATION OF THE COUPLED EQUATIONS OF MOTION FOR THE REDUCED DENSITY MATRIX FOR THE SAME BATH CASE

In this section we present the derivation of coupled equation of motion for the same bath case i.e. Eq. (13) from Eq. (12). The first step is the expansion of the reduced density matrix $\sigma$ in the eigenstates $\left|b_m\right\rangle$ of the bath diffusion operator $\Gamma$ (subscript is dropped), expressed as,

$$\sigma = \sum_m \sigma_m \left|b_m\right\rangle . \qquad \text{(A1)}$$

Eigen-value of the diffusion operator $\Gamma$ can be written as follows,

$$\Gamma\left|b_m\right\rangle = E_m\left|b_m\right\rangle . \tag{A2}$$

Substituting Eq. (A1) and Eq. (A2) into Eq. (12) we obtain (considering $\hbar = 1$)

$$\sum_m \frac{d\sigma_m}{dt}\left|b_m\right\rangle = -iH_{tot}^x \sum_m \sigma_m\left|b_m\right\rangle + \sum_m E_m\sigma_m\left|b_m\right\rangle . \tag{A3}$$

After that replacing the full Hamiltonian (Eq. (3) and Eq. (5)) into Eq. (A3) and followed by the multiplication with $\left\langle b_{m'}\right|$ we obtain

$$\frac{d\sigma_m}{dt} = -iH_s^x\sigma_m - i\sum_{m'}\left\langle b_{m'}\right|V_d(t)\left|b_m\right\rangle\left(\sum_k |k\rangle\langle k|\right)^x \sigma_{m'} - i\sum_{m'}\left\langle b_{m'}\right|V_{od}(t)\left|b_m\right\rangle\left(\sum_{\substack{k,l\\k\neq l}} |k\rangle\langle l|\right)^x \sigma_{m'} + E_m\sigma_m . \tag{A4}$$

For Poisson bath $\Gamma$ can be expressed as $2\times 2$ matrix as follows,

$$\Gamma = \begin{pmatrix} -\frac{b}{2} & \frac{b}{2} \\ \frac{b}{2} & -\frac{b}{2} \end{pmatrix}. \tag{A5}$$

Eigen-values of $\Gamma$ are $0$ and $-b$. The diagonal fluctuation $V_d(t)$ and off-diagonal fluctuation $V_{od}(t)$ have the following matrix elements in the eigen-functions of $\Gamma$ as follows

$\left\langle b_i\right|V_a(t)\left|b_j\right\rangle = V_a$, where $i \neq j$

$$\left\langle b_i\right|V_a(t)\left|b_i\right\rangle = 0 . \tag{A6}$$

$a$ in $V_a$ represents strength of either diagonal or off-diagonal fluctuation. Finally using the eigen-values and Eq. (A6) we obtain Eq. (13).

# APPENDIX B: DERIVATION OF THE COUPLED EQUATIONS OF MOTION FOR THE REDUCED DENSITY MATRIX FOR THE INDEPENDENT BATH CASE

For independent bath case from Eq. (15), following the same steps described for same bath case one can obtain Eq. (16). We provide a glimpse of the steps. For two sites model QSLE can be written as follows (using Eq. (15))

$$\frac{d\sigma}{dt} = -\frac{i}{\hbar}\left[H(t),\sigma\right] + \Gamma_1\sigma + \Gamma_2\sigma + \Gamma_3\sigma \tag{B1}$$

where $\Gamma_1$, $\Gamma_2$ and $\Gamma_3$ are the stochastic diffusion operator for $V_{11}$, $V_{22}$ and $V_{12}\left(=V_{21}\right)$ (two diagonal and one off-diagonal elements) respectively.

Next step is the expansion of the reduced density matrix in eigenfunctions of $\Gamma_1$, $\Gamma_2$ and $\Gamma_3$ as,

$$\sigma = \sum_{j,k,l} \sigma_{jkl} \left|b_j^1\right\rangle \left|b_k^2\right\rangle \left|b_l^3\right\rangle . \tag{B2}$$

where $\left|b_j^1\right\rangle$, $\left|b_k^2\right\rangle$ and $\left|b_l^3\right\rangle$ represent the $j^{th}$ bath state of the diffusion operator $\Gamma_1$, $k^{th}$ bath state of the diffusion operator $\Gamma_2$ and $l^{th}$ bath state of the diffusion operator $\Gamma_3$ respectively. Now one can make use of the same procedure described for same bath case to obtain coupled equation of motion for independent bath case (two sites model) i.e. Eq. (16). One can extrapolate the derivation of two sites model to obtain the coupled equation of motion for larger sites model.

# APPENDIX C: COUPLED EQUATION OF MOTION FROM RATE EQUATION DESCRIPTION

Coupled equation of motion from rate equation description for 7 site discrete model is provided as follows,

$$\frac{dP_1}{dt} = -kP_1 + kP_2 \tag{C1}$$

$$\frac{dP_2}{dt} = kP_1 - 2kP_2 + kP_3 \tag{C2}$$

$$\frac{dP_3}{dt} = kP_2 - 2kP_3 + kP_4 \tag{C3}$$

$$\frac{dP_4}{dt} = kP_3 - 2kP_4 + kP_5 \tag{C4}$$

$$\frac{dP_5}{dt} = kP_4 - 2kP_5 + kP_6 \tag{C5}$$

$$\frac{dP_6}{dt} = kP_5 - 2kP_6 + kP_7 \tag{C6}$$

$$\frac{dP_7}{dt} = kP_6 - kP_7 \tag{C7}$$

where $P_1$, $P_2$…. are the populations of site 1, site 2… respectively and $k$ is the rate constant for population transfer process. The above coupled differential equations can be solved numerically using Runge-Kutta fourth order method.